# Developing a UXR Point of View for Cognitive Accessibility in Mobile Learning with Generative AI

Fatima Ahmad Muazu *

School of Computing and Engineering, Bournemouth University Poole, UK, fahmad@bournemouth.ac.uk

Festus Adedoyin

School of Computing and Engineering, Bournemouth University Poole, UK, fadedoyin@bournemouth.ac.uk

Huseyin Dogan

School of Computing and Engineering, Bournemouth University Poole, UK, hdogan@bournemouth.ac.uk

Abiodun Adedeji

School of Computing and Engineering, Bournemouth University Poole, UK, adedejia@bournemouth.ac.uk

Melike Akca

School of Computing and Engineering, Bournemouth University Poole, UK, makca@bournemouth.ac.uk

Olumuyiwa Ayorinde

School of Computing and Engineering, Bournemouth University Poole, UK, ayorindeo@bournemouth.ac.uk

This study investigates how UX research (UXR) principles, combined with Large Language Model (LLM)-supported analysis, can be used to improve the quality of requirements for mobile learning systems designed for learners with cognitive disabilities. Using the UXR Point-of-View (PoV) pyramid as a methodological framework, the study progressed through four stages: foundational structuring of psychological, behavioral, and design layers; structured validation using the DeLone and McLean Information Systems Success Model and Quality Function Deployment (QFD); insight consolidation through the development of nine Cognitive Accessibility UXR Play Cards; and stakeholder-specific PoV articulation to support interdisciplinary communication. LLM-supported synthesis was integrated to assist in theme clustering, requirement refinement, and hypothesis formulation under human oversight. Findings suggest that many usability and engagement challenges in mobile learning originate from ambiguous or under-specified requirements rather than interface design alone. By embedding cognitive accessibility principles into measurable and technically traceable requirements, the proposed Cognitive Accessibility UXR Playbook provides a structured pathway for aligning theory, system architecture, and stakeholder strategy.

---

* Place the footnote text for the author (if applicable) here.

## 1 INTRODUCTION

The rapid growth of mobile learning has transformed access to education as it offers flexible and personalized learning opportunities across diverse contexts. However, while mobile learning applications are widely adopted, their effectiveness for learners with cognitive disabilities remains inconsistent. Previous research has identified determinants of mobile learning quality such as system, interaction and content quality [1], emotional engagement [2], adaptive learning styles [3], [4], and motivational design [5]. Limited attention has been given to how requirement quality directly shapes user experience (UX) outcomes for cognitively diverse learners.

For learners with cognitive disabilities, these factors are even more critical, because poorly designed interfaces may increase cognitive load, reduce motivation, and hinder knowledge retention. Cognitive disabilities, including conditions affecting memory, attention, processing speed and executive functioning, present unique challenges that require carefully designed learning instructions and user interfaces in mobile learning. The quality of mobile learning, therefore, cannot be assessed solely through technological performance or content availability rather, it must be examined through a comprehensive user experience (UX) perspective that prioritizes accessibility, usability, emotional engagement and cognitive load based on user specified requirements.

In the context of this study, quality function deployment (QFD) as applied on online learning [6] will be employed to assist in prioritizing technical features, it is a structured approach that transforms user requirements (voice of the user) into quality technical design specifications for mobile learning applications. The Information Systems success model particularly DeLone and McLean [7] will be applied to determine the quality of user requirements and predict mobile learning success, impliedly, user experience.

Existing UX approaches often provide high-level design principles without operational mechanisms for translating interdisciplinary knowledge into measurable and testable technical requirements. Furthermore, accessibility guidelines such as [8]'s cognitive accessibility guidelines are frequently treated as compliance checklists instead of being integrated into systemic requirement engineering processes. As AI-driven technologies becomes embedded within educational tools [9], the need for explainable, cognitively accessible and quality-driven design becomes even more critical.

While existing UX frameworks have emphasized system quality, interaction quality and content quality in mobile learning contexts [1], and user-centered methodologies such as mLUX have highlighted emotional and experiential factors [2], these approaches mostly remain broadly descriptive. Similarly, adaptive and personalized models in mobile learning [3,4,10] demonstrate the importance of learner-specific considerations yet provide limited procedural guidance for translating accessibility principles into structured requirement specifications. Studies focusing on special user groups, such as deaf learners [11], further reveal the necessity of targeted UX evaluation instruments; however, a comprehensive and systemic framework tailored specifically to cognitive accessibility remains underdeveloped.

Existing UXR research shows the imminent role of User Experience Research (UXR) frameworks and artificial intelligence in supporting transparent, evidence-based design decisions. Integrating Generative AI within the UXR Point of View (PoV) framework demonstrates how AI can support practitioners to break down complex research into structured and actionable design insights. [12] argue the lack of systematic methods that transform empirical evidence into perspectives that drive impact therefore proposed a UXR playbook that supported the formation of evidence-based viewpoints that improve decision making situational awareness. Example application of UXR playbook [13] addresses the challenges of designing Explainable Artificial Intelligence (XAI) interfaces and offers practical guidance to bridge technical explainability and user-centered design in order to foster trust, transparency and responsible AI adoption.

This research therefore investigates how UX research principles, combined with LLM-supported analysis, can be employed to determine and improve the quality of mobile learning applications designed for learners with cognitive disabilities. Emphasis is placed on the relationship between requirement quality and user experience outcomes, examining how well-defined, measurable and accessibility-oriented requirements can positively influence usability, engagement and learning effectiveness. The research proposes a UXR Point-of-View (PoV) methodology and a Cognitive Accessibility UXR Playbook that bridge technical explainability, interdisciplinary expertise and user-centered design. The framework integrates Information Systems success theory Delone and Mclean [7], Quality Function Deployment (QFD), cognitive accessibility guidelines and AI-supported requirement analysis to transforms insights from psychologists, educators and accessibility specialists into structured quality requirements for mobile learning systems tailored to cognitive disabilities.

## 2 METHODOLOGY

This study employed a mixed-method, AI-augmented User Experience Research (UXR) synthesis using the UXR Point-of-View (PoV) framework to evaluate mobile learning quality for learners with cognitive disabilities. The focus was on how the quality of requirements particularly, accessibility, clarity, adaptability and cognitive load considerations affect user experience and learning effectiveness.

At the Foundational level, accessibility principles were established through systematic literature review, stakeholder requirements and W3C cognitive accessibility guidelines. The Data level integrated qualitative feedback, requirement artefacts and evaluation models (e.g., DeLone and McLean IS model, QFD) to assess usability and system quality. The Insight level synthesized recurring cognitive accessibility patterns into actionable UX themes. Finally, the UXR PoV level articulated stakeholder-oriented narratives linking requirement quality to usability, trust, engagement and learning outcomes.

Generative AI supported thematic synthesis and insight refinement across all stages, while interpretative control remained with the research team.

### 2.1 Stage 1: leveraging GenAI and UXR PoV framework

At the foundational stage of the UXR PoV pyramid, the study clarified the strategic challenge, identified stakeholder ecosystems, defined success constructs, and established a structured requirement-validation workflow that anchored cognitive accessibility. The following prompts were used in ChatGPT.

- "Considering that the effectiveness of an application would rely on the quality of its requirement, relate these sets of requirements to Information system success model constructs."
- "Categorize the themes into the UXR Point of View Framework (psychological → behavioral → design layers)."
- "Considering that the effectiveness of an application would rely on the quality of its requirement, relate these sets of requirements to Information system success model constructs."
- "Generate one hypothesis per theme linking them to design features relying on UXR PoV pyramid described, the UXR PoV website."

### 2.2 Stage 2: Establishing Foundational Plan and Roadmap

In this stage of the UXR PoV pyramid, the study operationalized structured validation mechanisms to ensure requirement quality and system alignment. The DeLone and McLean Information Systems Success Model was applied to map emerging requirements against system quality, information quality, service quality, user satisfaction, use, and net benefits.

Based on systematic review [14], focus group dataset, and relevant accessibility guidelines, explain what barriers could hinder learning for learners with cognitive disabilities. This stage was achieved using the following prompts.

- “Formulate these barriers into a set of mobile learning design requirements and considering that the effectiveness of an application would rely on the quality of its requirement, relate these sets of requirements to Information system success model constructs.”
- “Leveraging on quality function deployment technique, map out the relationship between these requirements to appropriate technical requirements for cognitive accessibility in mobile learning in order of priority.”
- “Identify the key stakeholders for a project that effectively delivers cognitive accessibility in mobile learning and specify their roles, contributions and data needs.”

### 2.3 Stage 3: insights generation and play card development

This stage focused on insight consolidation through the generation of nine UXR Play Cards. Each card articulated a clearly defined issue, associated risks, evidence-based best practices, and corresponding design implications. By structuring insights into reusable decision-support artifacts, the study transformed thematic findings into actionable governance tools. The following prompts were used.

- “Create a step-by-step project plan linking research goal, user needs, and stakeholder engagement following a mixed-method approach (quantitative, qualitative, AI synthesis) emphasizing cognitive accessibility in mobile learning.”
- “Generate hypotheses play cards that links theory, evidence and design rationale that are in the UXR Playbook Cards structure.”
- “Translate identified technical requirements into design features and then structure into UXR Playbook Cards.”

### 2.4 Stage 4: PoV narratives and stakeholder communication

The final stage of the UXR PoV pyramid involved articulating stakeholder-specific Points of View that translated structured insights into strategic narratives. PoVs were adapted for executives, engineers, designers, educators, and policymakers. Each narrative highlighted core insights, key technical requirements, and the role of GenAI-supported analysis. This ensured that cognitive accessibility considerations were communicated in discipline-relevant terms, fostering alignment, adoption, and actionable decision-making. This was based on the following prompts.

- “Using the UXR PoV template provided, generate disability specific PoV narratives for dyslexia, ADHD and Autism and different stakeholders.”
- “Refine the PoV statement with technical explainability and user experience clarity.”
- “Summarize insights from Play Cards into concise stakeholder messages with emphasis on key technical requirements and implementation guidelines.”

## 3 RESULTS

### 3.1 Evidence Synthesis and Hypothesis Generation

The study structured the problem space using a three-tier UXR PoV analytical model comprising psychological, behavioral, and design layers in Figure 1.

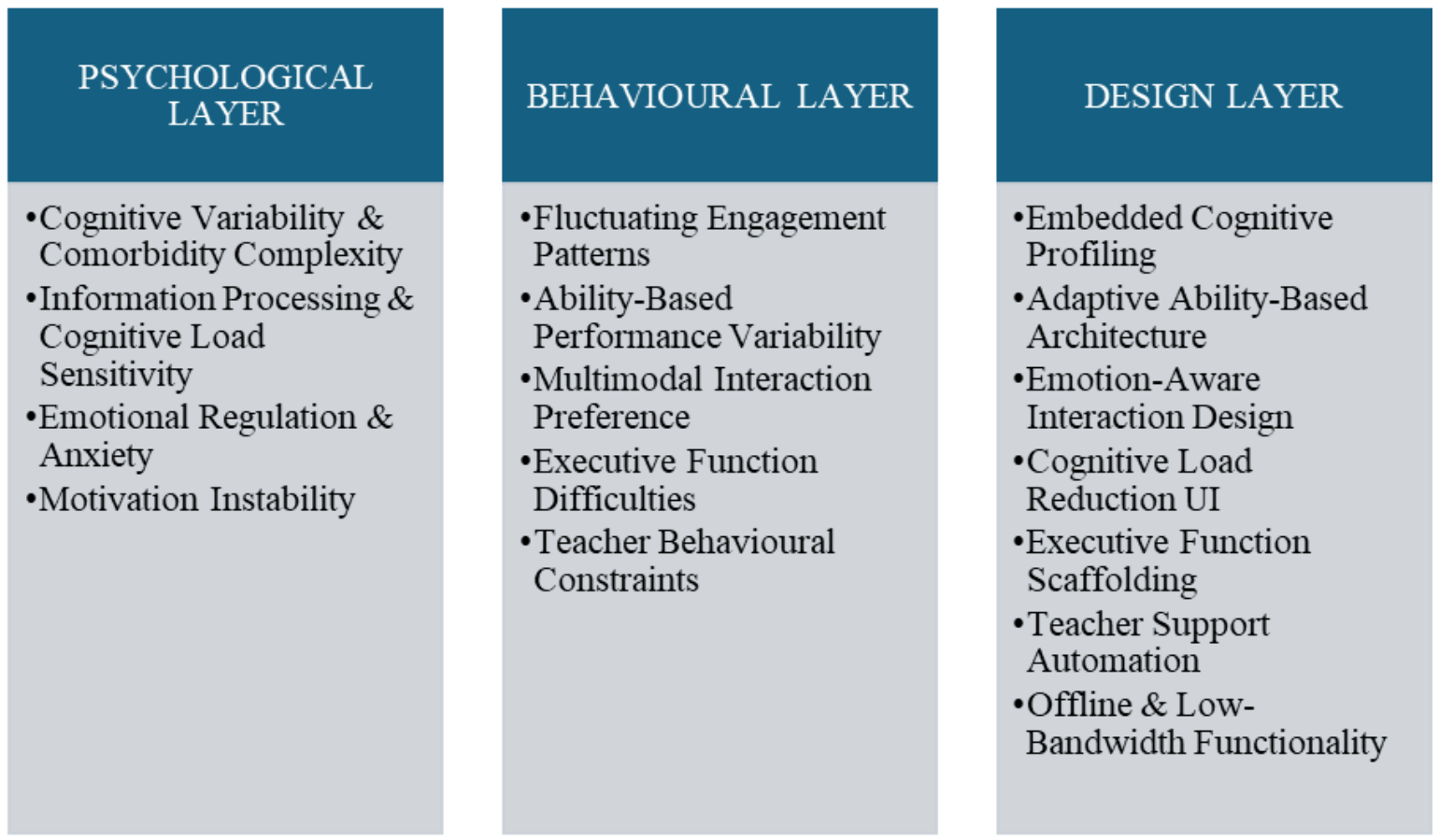

Figure 1: UXR PoV Foundational Structuring of themes

**Hypothesis**

***H1****: If the platform implements adaptive difficulty and step-by-step task segmentation, then learners will complete more tasks independently.*

***H2****: If the platform includes mood check-ins and a low-stimulation interface mode, then session duration and re-engagement rates will increase.*

***H3****: If engagement formats rotate based on learner interaction patterns, then engagement drop-off across sessions will decrease.*

***H4:*** *If learners can switch between visual, auditory, and interactive content formats, then comprehension accuracy will improve.*

***H5****: If the system auto-generates differentiated learning pathways, then teacher preparation time will decrease while personalization increases.*

***H6****: If the platform operates offline with low-bandwidth optimization, then adoption and sustained use in low-resource contexts will increase.*

***H7***: *If the platform provides embedded task planning and reminder scaffolds, then task initiation and completion consistency will increase.*

***H8:*** *If the system integrates ongoing cognitive profiling to adjust learning pathways, then ability–task alignment accuracy will improve.*

***H9***: *If the platform maintains a consistent, low-clutter interface layout, then navigation errors and user confusion will decrease.*

### 3.2 Foundational Plan and Stakeholder Roadmap

This layer ensured theoretical grounding and established a stakeholder mapping and further clarified interdisciplinary roles and adoption pathways ecosystem as presented in Table 1.

Table 1: Stakeholder Ecosystem Model

| Stakeholder | Primary Focus |
|---|---|
| Learners | Usability & engagement |

| Stakeholder | Primary Focus |
|---|---|
| Teachers | Personalization & scalability |
| Psychologists | Cognitive validity |
| Parents | Contextual behaviour data |
| UX researchers | Insight validation |
| Instructional designers | Pedagogical structure |
| Engineers | Technical feasibility |
| Accessibility experts | Compliance and inclusion |
| Parents | Contextual behaviour data |
| Administrators | Adoption & ROI |
| Policy Makers | Equity & regulation |
| Ethic officers | Data protection |

### *3.2.1 Integrating QFD*

After the application of DeLone and McLean IS model to emerging requirements, QFD was then used to translate accessibility-oriented requirements into prioritized technical specifications in figure 2, establishing traceability between user needs and engineering implementation. Validated cognitive accessibility requirements are mapped to corresponding technical requirements in order of priority in Table 2. Strength of relationship is Strong = 9, Moderate = 3, Weak = 1.

TR1 • Real-time performance analytics engine
TR2 • Adaptive learning algorithm (rule-based or ML)
TR3 • Modular content architecture (micro-learning object model)
TR4 • State-aware UI system (emotion-sensitive rendering logic)
TR5 • Low-cognitive-load UI framework (consistent layout templates)
TR6 • Multimodal rendering engine (TTS, STT, captions, icons)
TR7 • Engagement tracking & reinforcement engine
TR8 • Automated IEP & pathway generation engine
TR9 • Offline-first data storage & sync architecture
TR10 • Low-bandwidth optimization layer
TR11 • Accessibility compliance layer (WCAG/WAI cognitive patterns)

Figure 2: Technical requirements

Table 2: QFD Relationship Matrix

| Customer Requirements (WHATs) | TR1 | TR2 | TR3 | TR4 | TR5 | TR6 | TR7 | TR8 | TR9 | TR10 | TR11 |
|---|---|---|---|---|---|---|---|---|---|---|---|
| CR1 Adaptive difficulty | 9 | 9 | 9 | 1 | 3 | 1 | 3 | 3 | 1 | 1 | 3 |
| CR2 Emotional support | 3 | 3 | 1 | 9 | 9 | 3 | 3 | 1 | 1 | 1 | 9 |
| CR3 Ability-based progression | 9 | 9 | 9 | 1 | 3 | 1 | 1 | 9 | 1 | 1 | 3 |
| CR4 Simple interface | 1 | 1 | 3 | 3 | 9 | 3 | 1 | 1 | 1 | 1 | 9 |
| CR5 Multimodal delivery | 1 | 1 | 3 | 3 | 3 | 9 | 3 | 1 | 1 | 1 | 9 |
| CR6 Engagement management | 9 | 3 | 1 | 3 | 3 | 3 | 9 | 3 | 1 | 1 | 3 |
| CR7 Teacher automation | 9 | 3 | 3 | 1 | 1 | 1 | 3 | 9 | 1 | 1 | 3 |
| CR8 Offline accessibility | 1 | 1 | 3 | 1 | 3 | 1 | 1 | 1 | 9 | 9 | 3 |

## 3.3 UXR play cards

The play Cards formalized a Cognitive Accessibility UXR Playbook that bridged theory, empirical evidence, and technical rationale. Thematic findings were synthesized into nine structured UXR Play Cards, each detailing a specific issue, its systemic risk, recommended best practice, and design rationale. Figure 3 illustrates front and back samples of play cards for adaptive difficulty such as cognitive scaffolding and multimodal flexibility.

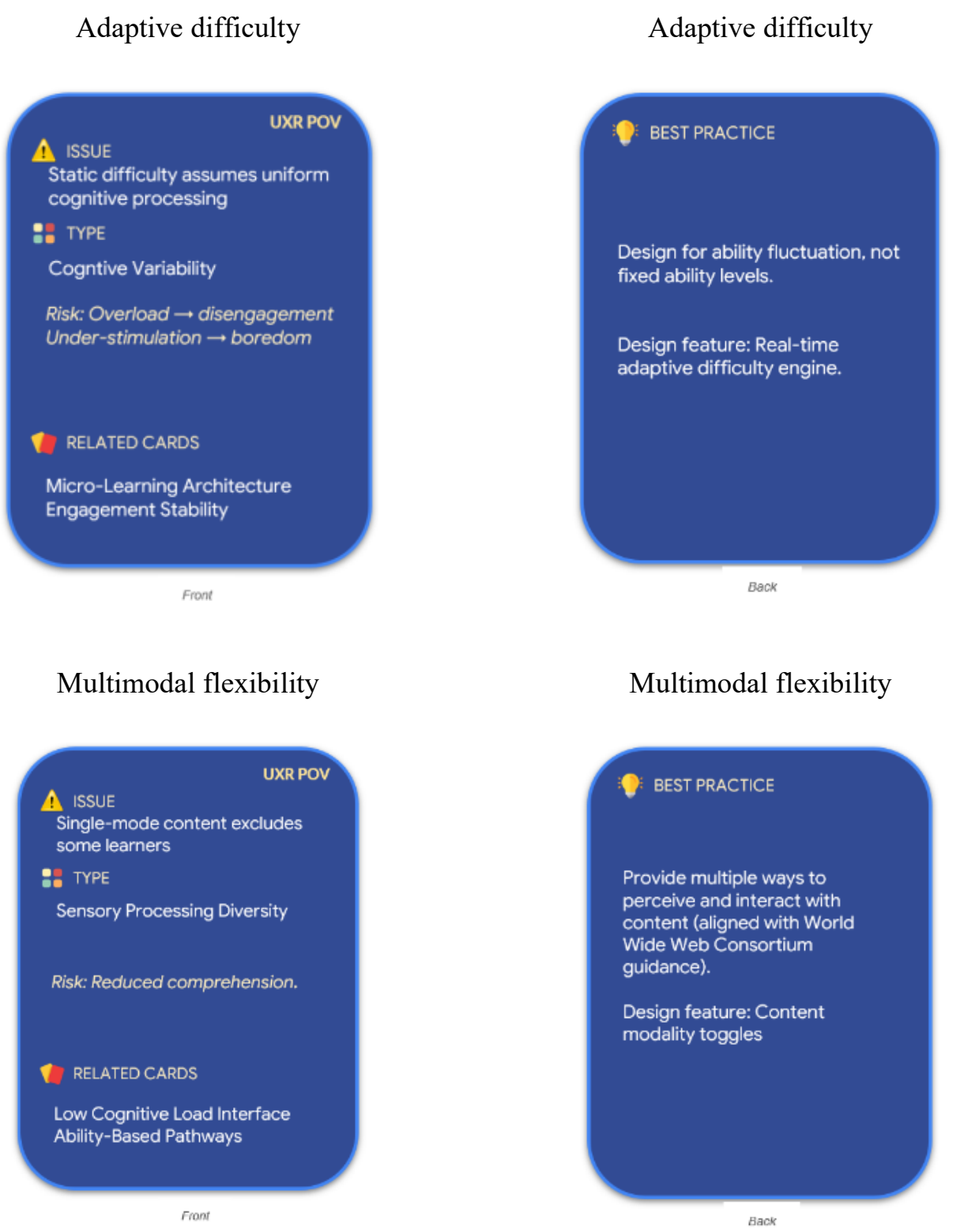

Figure 3: Adaptive Difficulty as Cognitive Scaffolding and Multimodal flexibility.

## 3.4 PoV Narratives.

This ensured that cognitive accessibility considerations were communicated in discipline-relevant terms, fostering alignment, adoption, and actionable decision-making. Different perspectives were crafted for executive, technical, design, educational, and policy audiences, each integrating core insights, priority technical requirements, and GenAI-supported contributions. Figure 4 is a snapshot of PoV narrative for UX designers.

| Stakeholder | Core Insight | Key Technical Requirements | PoV Narrative Focus | GenAI-Supported Contribution |
|---|---|---|---|---|
| Designers/UX Team | Cluttered, unpredictable interfaces increase anxiety and reduce cognitive comprehension | Consistent, Simplified Interaction Framework | Cognitive accessibility in mobile learning demands predictable, simplified interaction patterns that reduce extraneous cognitive load. | -Simulate cognitive load scenarios.<br>-Generate alternative layout variants for testing<br>-Auto-evaluate UI copy complexity.<br>-Assist in rapid accessibility audit generation. |

Figure 4: Stakeholder Specific Narrative

## 4 DISCUSSION & CONCLUSION

This study advances mobile learning research by repositioning requirement quality as a primary determinant of user experience outcomes for learners with cognitive disabilities. While prior studies have examined UX components, motivation strategies, adaptive systems, and emotion-aware models, few have interrogated the upstream question: *How well are the system requirements defined, structured, and validated before design and implementation?*

The UXR PoV integration ensured that these adaptive elements were not treated as isolated features. Instead, they were derived from structured insights and encoded as measurable requirements (e.g., adaptive difficulty engines, micro-learning modularity, emotion-aware UI logic). This distinction is critical: adaptation becomes an architectural commitment grounded in UX-derived evidence rather than a surface-level enhancement. The integration of LLM-supported analysis aligns with the AI-augmented UXR vision described in the CHI workshop [15], the LLMs functioned as augmentation tools within a human-in-the-loop process. They enhanced efficiency in synthesis and cross-mapping tasks but did not replace interpretative judgement. This reflects responsible AI principles and mirrors calls within XAI literature [13] to bridge technical complexity with user-centred understanding.

The findings suggest that many UX failures in mobile learning originate in under-specified requirements rather than interface design alone. By integrating UXR PoV methodology with structured requirement engineering and GenAI-supported synthesis, this research provides a repeatable framework for designing cognitively accessible mobile learning systems that are both inclusive and outcome-driven.

## REFERENCES

[1] T. Jiang, G. Luo, Z. Wang, and W. Yu, “Research into influencing factors in user experiences of university mobile libraries based on mobile learning mode,” *Libr. Hi Tech*, vol. 42, no. 2, pp. 564–579, May 2024.

[2] A. Dirin and M. Nieminen, “mLUX: Usability and User Experience Development Framework for M-Learning,” *Int. J. Interact. Mob. Technol.*, vol. 9, no. 3, p. 37, Jul. 2015.

[3] A. Nimkoompai and W. Paireekreng, “Dynamic UX based m-learning using user profile of learning style,” in *Proceedings of the 3rd International Conference on Communication and Information Processing*, 2017, pp. 221–225.

[4] A. Nimkoompai and W. Paireekreng, “Framework for UX Based M-Learning Using Learning Style and Recommendation System,” *Int. J. Informatics Inf. Syst.*, vol. 3, no. 3, pp. 106–113, 2020.

[5] J. Seppala, T. Mitsuishi, Y. Ohkawa, X. Zhao, and M. Nieminen, “Study on UX design in enhancing student motivations in mobile language learning,” in *2020 IEEE International Conference on Teaching, Assessment, and Learning for Engineering (TALE)*, 2020, pp. 948–951.

[6] J. R. Al yaarubi and A. Rajakannu, “Implementation of Quality Function Deployment to Improve Online Learning and Teaching in Higher Education Institutes of Engineering in Oman,” *Int. J. Learn. Teach. Educ. Res.*, vol. 23, no. 12, pp. 463–486, Dec. 2024.

[7] W. H. DeLone and E. R. McLean, “The DeLone and McLean Model of Information Systems Success: A Ten-Year Update,” *J. Manag. Inf.*

*Syst.*, vol. 19, no. 4, pp. 9–30, 2003.

[8] WAI, “W3C Web Accssibility Inititaitve,” *Cognitive Accessibility at W3C*, 2019. [Online]. Available: https://www.w3.org/WAI/cognitive/. [Accessed: 15-Jul-2025].

[9] O. J. F. Chavez and T. Palaoag, “UI/UX prototype design for a personalized learning mobile app to boost comprehension: a design thinking model,” *TQM J.*, Jun. 2025.

[10] L. Shen, B. Xie, and R. Shen, “Enhancing User Experience in Mobile Learning by Affective Interaction,” in *2014 International Conference on Intelligent Environments*, 2014, pp. 297–301.

[11] N. Mohamad and N. L. Hashim, “UX Testing for Mobile Learning Applications of Deaf Children,” *Int. J. Adv. Comput. Sci. Appl.*, vol. 12, no. 11, 2021.

[12] H. Dogan, S. Giff, and R. Barsoum, “User Experience Research: Point of View Playbook,” in *Extended Abstracts of the CHI Conference on Human Factors in Computing Systems*, 2024, pp. 1–7.

[13] M. Naiseh, H. Dogan, S. Giff, and N. Jiang, “Development of a persuasive User Experience Research (UXR) Point of View for Explainable Artificial Intelligence (XAI),” in *The workshop held at the ACM Conference on Human Factors in Computing Systems*, 2025.

[14] F. A. Muazu, F. A. Adedoyin, H. Dogan, N. Mavengere, and P. Whittington, “The use of mobile learning in special education needs and disabilities (SEND) settings: state-of-the-art classification of studies,” *ACM Int. Conf. Proceeding Ser.*, pp. 486–495, 2024.

[15] H. Dogan, R. M. Barsoum, S. Giff, A. Dix, and E. Churchill, “Defining a UX Research Point of View (POV),” in *Proceedings of the Extended Abstracts of the CHI Conference on Human Factors in Computing Systems*, 2025, pp. 1–3.